\documentclass[showpacs,aps]{revtex4}
\usepackage{dcolumn}
\usepackage{amssymb}
\usepackage{color}
\usepackage[colorlinks=true,linkcolor=red]{hyperref}
\usepackage{hyperref}
\begin{document}
\title{Fermions on the brane in 6D with nonsingular exponential scale factors}
\author{Pavle  Midodashvili}
\email{midodashvili@hotmail.com} \affiliation{Tskhinvali State
University, 2 Besiki Str., Gori 383500, GEORGIA}

\date{\today}
\begin{abstract}We introduce new realistic brane-solutions with exponential scale
factors in the $6D$-space-time. We show that for these solutions
the zero modes of all bulk fields are sharply localized at
different positions on the brane and have "Gaussian shape"
wave-functions in the extra space. We also explicitly show that in
the model there are cases when exactly three fermion generations
naturally arise only through gravity. Because of localized fermion
modes are also stuck at different positions in the extra space,
there is possibility to provide a framework for natural explaining
the fermion mass hierarchy in terms of higher dimensional
geography.\end{abstract}

 \maketitle

\section{Introduction}
The idea of extra dimensions \cite{R-Sh, Akama,Visser,G-W,Ant,R-S}
is one of the most attractive ideas concerning unification of
gauge fields with general relativity and new solutions to old
problems (smallness of cosmological constant, the origin of the
hierarchy problem, the nature of flavor, etc.). In theories with
extra dimensions our world is associated with a brane, embedded in
a higher-dimensional space-time with non-compact extra dimensions
and non-factorizable geometry. The key ingredient for realizing
the brane world idea is localization of various bulk fields on a
brane by a natural mechanism. The gravity is known to be the
unique interaction having universal coupling with all matter
fields, so it is important to find a purely gravitational trapping
mechanism. The brane solutions and matter localization mechanisms
has been widely investigated in scientific literature
\cite{AH-Dim-Dv,Ant-AH-Dim-Dv,Ch-Pop,Coh-Kap,Gh-Sh,B-G,Pomarol,Gregory,
Ch-Nel,Oda,Oda2,Gh-R-Sh,K-M-Ol,RD-Sh,Gog,mg1,mg2,Paul1}. Recently
the gravitational trapping of the zero modes of all bulk fields
was realized for the brane solutions with non-exponential warp
factors in $6$-dimensional bulk space-time
\cite{Paul2,GogberSingleton,Oda1}. In our previous article
\cite{Paul3} in the case of the non-exponential scale factors we
also have suggested new purely gravitational mechanism explaining
the origin of three generations of the Standard Model fermions and
explicitly show that localized fermions are stuck at different
points on the brane in the extra space. In these models, with
non-exponential scale factors, the zero-mode solutions of the bulk
fields are normalizable and integrals of there Lagrangians over
the extra coordinates are finite, but their wave functions spread
rather widely in the bulk owing to the lack of the non-exponential
warp factor. Thus, in order not to contradict the strict
experiments such as the charge conservation law, some parameters
in this models must be chosen in a proper way. In some cases it is
hard to say whether there is such a suitable choice of the
parameters. From this point of view it is interesting to consider
the models with exponential warp factors when the wave functions
of localized fields are sharply peaked on the brane in the extra
space. At present it is known (see for example \cite{B-G,Oda2})
that in $5D$ and $6D$ models with exponential warp factors it is
possible to localize all spin fields if one introduces
non-gravitational interactions. So it is of great interest to
realize a purely gravitational localization mechanism in the
models with exponential scale factors.

In this article in the case of $(1+5)$-space-time we introduce new
brane-solutions with exponential scale factors sharply localizing
the zero modes of all local fields (spin $0$ scalar field, spin
${\raise0.7ex\hbox{$1$} \!\mathord{\left/
 {\vphantom {1 2}}\right.\kern-\nulldelimiterspace}
\!\lower0.7ex\hbox{$2$}}$ spinor field, spin 1 gauge field, spin
${\raise0.7ex\hbox{$3$} \!\mathord{\left/
 {\vphantom {3 2}}\right.\kern-\nulldelimiterspace}
\!\lower0.7ex\hbox{$2$}}$ gravitino field, spin $2$ gravitational
field and totally antisymmetric tensor fields) on the brane in the
extra space, and for the realistic physical situation of matter
distribution in the $6D$-space-time we realize the purely
gravitational trapping mechanism explaining the origin of three
generations of the Standard Model fermions from one generation in
a higher-dimensional theory and explicitly show that fermions are
sharply stuck at different points on the brane in the extra space.

\section{ Main notations and equations}
The action and Einstein's equations which we consider in this
article have the form:\begin{equation}\label{MainAction}S = \int
{d^6 x\sqrt { - g} \left[ {\frac{{M^{4} }}{2}\left( {R + 2\Lambda
} \right) + L} \right]}~,~~R_{AB}  - \frac{1}{2}g_{AB} R =
\frac{1}{{M^{4}}}\left( {\Lambda g_{AB}  + T_{AB} }
\right)~,\end{equation} where $M$, $R$, $\Lambda$, $L$, $R_{AB}$
and $T_{AB}$ are respectively the fundamental scale, the scalar
curvature, the cosmological constant, the Lagrangian of matter
fields, the Ricci and the energy-momentum tensors. In the bulk
space-time we use the following metric ansatz
\cite{Paul1,Paul2,GogberSingleton,Oda1,Paul3} : $ds^2  = \phi ^2
\left( r \right)g _{\alpha \beta } dx^\alpha dx^\beta - g\left( r
\right)\left( {dr^2  + r^2 d\theta ^2 } \right)$. The source of
the brane is described by a stress-energy tensor $T_{AB}$, its
nonzero components we choose in the form $T_{\alpha \beta}   = -g
_{\alpha \beta} F_0 \left( r \right),\ \ T_{ij}  = -g_{ij} F\left(
r \right)$, where $F_{0}$ and $F$ are source functions, which
depend only on the radial coordinate $r$ in the extra space. In
this case the Einstein's equations take the following form
\begin{equation}\label{E1}\frac{{3\phi ''}}{\phi } + \frac{{3\phi '^2 }}{{\phi ^2 }} +
\frac{{3\phi '}}{{r\phi }} + \frac{{g''}}{{2g}} - \frac{{g'^2
}}{{2g^2 }} + \frac{{g'}}{{2rg}} = \frac{{g\left( {F_0  - \Lambda
} \right)}}{{M^4 }} + \frac{{g\Lambda _{phys} }}{{\phi ^2 M_P^2
}},\end{equation}
\begin{equation}\label{E2}\frac{{6\phi '^2 }}{{\phi ^2 }} + \frac{{2\phi 'g'}}{{\phi g}} +
\frac{{4\phi '}}{{r\phi }} = \frac{{g\left( {F - \Lambda }
\right)}}{{M^4 }} + \frac{{2g\Lambda _{phys} }}{{\phi ^2 M_P^2
}},\ \ \frac{{4\phi ''}}{\phi } + \frac{{6\phi '^2 }}{{\phi ^2 }}
- \frac{{2\phi 'g'}}{{\phi g}} = \frac{{g\left( {F - \Lambda }
\right)}}{{M^4 }} + \frac{{2g\Lambda _{phys} }}{{\phi ^2 M_P^2
}},\end{equation} where the prime denotes differentiation $d/dr$.
The constant $\Lambda _{phys}$ represents the physical
four-dimensional cosmological constant, where
 $R_{\alpha \beta
}^{(4)}  - \frac{1}{2}g_{\alpha \beta } R^{(4)} = \frac{{\Lambda
_{phys} }}{{M_P^2 }}g_{\alpha \beta }$. In this equation
$R_{\alpha \beta }^{(4)}$, $R^{(4)}$ and $M_P$ are
four-dimensional physical quantities: Ricci tensor, scalar
curvature and Planck scale.
\section{ Nonsingular brane solution with exponential scale factors}
In the case \ $\Lambda _{phys}=0$ \  from the equations
(\ref{E1})-(\ref{E2}) we can find \cite{mg1,mg2,Paul3} : $F' +
4\frac{{\phi '}}{ \phi }\left( {F - F_0 } \right) = 0\ ; \ \ \ g =
\frac{{\delta \phi '}}{r} \ ; \ \ \ \ r\frac{{\phi ''}}{\phi } +
3r\frac{{\phi '^2 }}{{\phi ^2 }} + \frac{{\phi '}}{\phi } =
\frac{{rg}}{{2M^4 }}\left( {F - \Lambda } \right)$, where $\delta$
denotes the integration constant with the units of squared length.
The possible two solutions to this equations are (in what follows
we call these solutions "internal solutions" and introduce
subscript "int"):
\begin{enumerate}
\item The first solution with scale factor $\phi_{{\mathop{\rm int}} }$ monotonically decreasing
from the value equal to $1$ at the origin $z=0$ in the extra space
and asymptotically approaching the the nonzero positive value
equal to $1-d$ at the radial infinity $z=+\infty$
\begin{equation}\label{internal1}\phi_{{\mathop{\rm int}} } \left( z \right) = 1 - d\left[ {1 - e^{ - z^p } }
\right],\ \ g_{{\mathop{\rm int}} }\left( z \right) = {\delta
dp}{\varepsilon ^{-2} }z^{p - 2} e^{ - z^p },\ \ 0 < d < 1,\ \ p
\ge 2,\ \delta > 0,\end{equation}
\item The second solution with scale factor $\phi_{{\mathop{\rm int}} }$ monotonically
increasing from the value equal to $1$ at the origin $z=0$ in the
extra space and asymptotically approaching the positive value
equal to $1+d$ at the radial infinity $z=+\infty$
\begin{equation}\label{internal2}\phi_{{\mathop{\rm int}} } \left( z \right) = 1 + d\left[ {1 - e^{ - z^p } }
\right],\ \ g_{{\mathop{\rm int}} }\left( r \right) ={\delta
dp}{\varepsilon ^{-2} } z^{p - 2} e^{ - z^p } ,\ \ d
>0, \ p \ge 2, \ \delta > 0.\end{equation}\end{enumerate} In
(\ref{internal1})-(\ref{internal2}) we have introduced the
dimensionless radial coordinate $z = {r \mathord{\left/ {\vphantom
{r \varepsilon }} \right . \kern-\nulldelimiterspace} \varepsilon
}$, where $\varepsilon$ is some positive constant with the units
of length. In both cases the scale factor $\phi_{{\mathop{\rm
int}} }$ has an inflection point $r_{\rm{infl.\phi}} = \varepsilon
\left( {\frac{{p - 1}}{p}} \right)^{\frac{1}{p}}$ (this value
characterizes the order of the "effective width" $\Delta
_{{\rm{Brane}}}$ of the "thick brane" in the extra space). These
solutions have equal scale functions $g_{{\mathop{\rm int}} .}$.
In the case $p=2$ this function monotonically decreases from value
equal to $\frac{{2\delta d}}{{\varepsilon ^2 }}$ at the origin
$z=0$ to the value equal to zero at the infinity $z=+\infty$ in
the extra space and its derivative $g'_{{\mathop{\rm int}} }$ is
equal to zero at the origin and at the infinity in the extra
space. In the case $p>2$ this function is equal to zero at the
origin $r=0$ in the extra space and monotonically increases in the
range
 $0 <r < r_{\max.g }$, approaches the maximum value at the $r_{\max.g}$ and then
monotonically decreases in the range $r_{\max.g} < r < +\infty$
and approaches the asymptotic value equal to zero at the radial
infinity, where  $r_{\max.g }  = \varepsilon \left( {\frac{{p -
2}}{p}} \right)^{\frac{1}{p}}$. It must be mentioned that the
scale factor $g_{{\mathop{\rm int}} }(r)$ defines the topology of
the brane. These solutions exist for any $\Lambda$ and
corresponding source functions $F_{0}$ and $F$ can be easily
calculated from  corresponding Einstein's equations using
(\ref{internal1}) and (\ref{internal2}): $F = \Lambda  -
\frac{{2pM^4 }}{\delta }\phi ^{ - 2} \left[ {\left( {4\phi  - 3c}
\right)\ln \left( {\frac{{\phi  - c}}{{1 - c}}} \right) + \phi }
\right]$, $F_0  = F + \frac{\phi }{4}\frac{{dF}}{{d\phi }}$, where
$c=1-d$.

Similarly, as it was done in our article \cite{Paul3} for the case
of non-exponential scale factors, it is easy to explicitly show
that these solutions localize the zero modes of the following bulk
local fields: spin $0$ scalar field, spin 1 gauge field, spin
${\raise0.7ex\hbox{$3$} \!\mathord{\left/ {\vphantom
{32}}\right.\kern-\nulldelimiterspace} \!\lower0.7ex\hbox{$2$}}$
gravitino field, spin $2$ gravitational field and totally
antisymmetric tensor fields. The zero modes of these fields will
be localized at different positions on the brane in the extra
space, but as against a former case \cite{Paul3}, in this case we
shall have sharp localization of the modes with "Gaussian shapes"
of wave-functions in the extra space due to exponential type of
scale factors. Here we want to consider in more details the
localization of the bulk spin-${\raise0.7ex\hbox{$1$}
\!\mathord{\left/ {\vphantom{12}}\right.\kern-\nulldelimiterspace}
\!\lower0.7ex\hbox{$2$}}$ fermion in the background solution
(\ref{internal1}) (for the solution (\ref{internal2}) the results
are the same). Of course, in due analysis, in what follows we will
neglect the back-reaction on the geometry induced by the existence
of the bulk fields, and, without loss of generality, we will take
a flat metric on the brane. Fermion action and corresponding
equation of motion are
\begin{equation}\label{FermionAction0}
S_{\frac{1}{2}}  = \int {d^6 x\sqrt { - {}^6g} \overline \Psi
i\Gamma ^M D_M \Psi } \  ;\ \ \Gamma ^M D_M \Psi  =  0 \ \
.\end{equation} Introducing the vielbein $h_A^{\widetilde A}$
through the usual definition $g_{AB}  = h_A^{\widetilde A}
h_B^{\widetilde B} \eta _{\widetilde A\widetilde B}$ where
$\widetilde A,\widetilde B,...$ denote the local Lorentz indices,
it is easy to find the  non-vanishing components of the
spin-connection for the background (\ref{internal1}): $\omega
_\alpha ^{\widetilde z\widetilde\alpha }  = \frac{{\phi
'}}{{\varepsilon\sqrt g }},\ \ \ \omega _\theta^{\widetilde
z\widetilde\theta} = \frac{{\partial _z \left( {z\sqrt g }
\right)}}{{\sqrt g }}$. Therefore the covariant derivatives have
form $D_\alpha  \Psi  = \left( {\partial _\alpha   +
\frac{1}{2}\omega _\alpha ^{\widetilde z\widetilde\alpha } \gamma
_z \gamma _\alpha } \right),\ \ D_z \Psi  = \partial _z \Psi ,\ \
D_\theta \Psi  = \left( {\partial _\theta  + \frac{1}{2}\omega
_\theta^{\widetilde z\widetilde\theta} \gamma _z \gamma _\theta }
\right)\ $. The $6$-dimensional spinor $\Psi \left( {x^M }
\right)$ can be decomposed on a $4$-component spinor $\psi \left(
{x^\mu  } \right)$ and $2$-component spinor $\xi  \left( z,\theta
\right)$ :  $~\Psi \left( {x^M } \right) = \psi \left( {x^\mu  }
\right)\otimes \xi  \left( z,\theta \right)$. We require that the
$4$-dimensional spinor satisfies the equation $\gamma ^\mu
\partial _\mu  \psi \left( {x^\beta  } \right) = m\psi \left( {x^\beta  } \right)$. The $2$-component spinor
can can be expanded over the states with fixed rotation momentum $l$
in the $\theta$ direction: $\xi  \left( z,\theta \right)=\sum {\xi
_l \left( z \right)e^{il\theta } }$, where $\xi _l \left( z
\right)=(\upsilon_l,\zeta_l)$ is $2$-component spinor, $\upsilon_l$
and $\zeta_l$ are functions of variable $z$. Taking gamma matrices
$\gamma ^z $ and $\gamma ^\theta$  in the form $~~\gamma ^z  =
\left( {\begin{array}{*{20}c}   0 & 1  \\   1 &
0\\\end{array}} \right)~~,~~\gamma ^\theta   = \left( {\begin{array}{*{20}c}0 & { - i}  \\   i & 0  \\
\end{array}} \right)$, as a result we
obtain the following equations for $\upsilon_l$ and $\zeta_l$
\begin{equation}\label{FermionEquationZeroModesGeneral}\begin{array}{c}
\frac{m\varepsilon}{\phi }\upsilon _l  + \frac{1}{{\sqrt
{g_{{\rm{int}}{\rm{}}}} }}\left[ {\partial _z  + 2\frac{{\phi
_{{\rm{int}} {\rm{}}} '}}{\phi _{{\rm{int}}{\rm{}}}} +
\frac{1}{2}\frac{{\partial _z \left(
{zg_{{\rm{int}}{\rm{}}}^{\frac{1}{2}} }
\right)}}{{zg_{{\rm{int}}{\rm{}}}^{\frac{1}{2}} }} +\frac{l}{z}}
\right]\zeta_{l} = 0~,\\
\frac{m\varepsilon}{\phi }\zeta _l  + \frac{1}{{\sqrt
{g_{{\rm{int}}{\rm{}}}} }}\left[ {\partial _z + 2\frac{{\phi
_{{\rm{int}} {\rm{}}} '}}{\phi _{{\rm{int}}{\rm{}}}} +
\frac{1}{2}\frac{{\partial _z \left(
{zg_{{\rm{int}}{\rm{}}}^{\frac{1}{2}} }
\right)}}{{zg_{{\rm{int}}{\rm{}}}^{\frac{1}{2}} }} - \frac{l}{z}}
\right]\upsilon_{l}= 0~.\\
\end{array}\end{equation} In the zero-mass mode case $m=0$ (the case $m\ne 0$ we will consider later on)
the solution to these equations reads $\upsilon_l \left( z \right)
= a_l \phi _{{\rm{int}}{\rm{}}}^{ - 2} g_{{\rm{int}}{\rm{}}}^{ -
\frac{1}{4}} z^{ - \frac{1}{2} + l}~ ,~\zeta_l \left( z \right) =
b_l \phi _{{\rm{int}}{\rm{}}}^{ - 2} g_{{\rm{int}}{\rm{}}}^{ -
\frac{1}{4}} z^{ - \frac{1}{2} - l}~,~\xi _l \left( z \right) =
\phi _{{\rm{int}}{\rm{}}}^{ - 2} g_{{\rm{int}}{\rm{}}}^{ -
\frac{1}{4}} z^{ - \frac{1}{2}}\left( {\begin{array}{*{20}c}{a_l
z^l}\\   {b _l z^{-l}}  \\\end{array}} \right)$, with $a_l$ and
$b_l$ being the integration constants. Using explicit form
(\ref{internal1}) of the scale functions $\phi _{{\rm{int}}}$ and
$g_{{\rm{int}}}$ , the normalization requirement becomes $1 = 2\pi
\left( {\delta dp\varepsilon ^2 } \right)^{\frac{1}{2}}$
$\int\limits_0^{ + \infty } {\left[ {a_l^2 z^{2l } + b_l^2 z^{- 2l
} } \right]z^{\frac{1}{2}p - 1} e^{ - \frac{1}{2}z^p } dz}$, and
it is easy to see that the extra part of the $6D$-wavefunction is
normalizable (i.e., the integral over $z$ converges) only in the
following cases: I) $a_l\ne0~,~b_l=0~,~l> -p/4$ ; II)
$a_l=0~,~b_l\ne0~,~l<p/4$ ; III) $a_l\ne0~,~b_l\ne0~,~-p/4<l<p/4$.
Finally for the effective wave function of the zero-mass mode of
$6$-dimensional fermion in flat space we have $\Psi  = \psi \left(
{x^\mu  } \right) \otimes \left( {\delta dp\varepsilon ^2 }
\right)^{\frac{1}{4}} z^{\frac{1}{4}p - \frac{1}{2}} e^{ -
\frac{1}{4}z^p } \left[ {\sum\limits_{ - \frac{p}{4} < l <
\frac{p}{4}} {\left( {\begin{array}{*{20}c}
{a_l z^l e^{il\theta } }  \\{b_l z^{ - l} e^{il\theta } }  \\
\end{array}} \right)} } \right.\left. { + \sum\limits_{l \ge \frac{p}{4}} {\left( {\begin{array}{*{20}c}
{a_l z^l e^{il\theta } }  \\{b_{-l} z^l e^{ - il\theta } }  \\
\end{array}} \right)} } \right]$.
In this case because of exponential decreasing warp factors
normalizability of wave function means sharp localization of
corresponding mode on the brane. These components of zero-mass
mode of $6$-dimensional bulk fermion are localized at different
points \ \ $r_{{\rm{max}}{\rm{.,l}}}  = \varepsilon \left(
{\frac{{p + 4l - 2}}{p}} \right)^{\frac{1}{p}}$ in the extra
space. Each component can be regarded as $4$-dimensional massless
fermion. So in this case there are infinite number of massless
fermions on the brane. This situation indicates that in order to
single out finite number of these localized fermion components we
need some new mechanism, and the one possible way we will consider
in the next section.

To conclude this section let us consider the solution to
Einstein's equations without any sources ($F_0 \left( r \right)
\equiv F\left( r \right) \equiv 0$) in the case $\Lambda \ne 0,\ \
\Lambda _{phys}=0$. In this case the Einstein's equations reduce
to $r\frac{{\phi ''}}{\phi } + 3r\frac{{\phi '^2 }}{{\phi ^2 }} +
\frac{{\phi '}}{\phi } = - \frac{{rg}}{{2M^4 }}\Lambda \ , \ \ g =
\frac{{\delta \phi '}}{r},\ \ \delta  = const$. The solution to
these equations is (in what follows we call it "external solution"
and introduce corresponding subscript "ext")
\begin{equation}\label{SolutionNoSource}\phi_{{\rm{ext}}} \left( z \right) =
\frac{1}{{a + b\ln z}},\ \ g_{{\rm{ext}}}\left( z \right) =
\frac{{\delta b}}{{\varepsilon ^2 }}\frac{1}{{z^2 \left[ {a + b\ln
z} \right]^2 }},\ \ b =   \frac{{\delta \Lambda }}{{10M^4
}},\end{equation} where the  $a$ is some new positive parameter
and the constants $\varepsilon$, $\delta$ and the radial
coordinate $z$ are the same as above. This solution is nonsingular
in the range $e^{ - \frac{a}{b}}  < z <  + \infty $ in the extra
space. Its scale factors $\phi_{{\rm{ext}}.}$ and
$g_{{\rm{ext}}.}$ are monotonically decreasing functions in this
range.
\section{Three fermion generations on the brane}
Now imagine the following realistic physical situation. Suppose in
the extra space in the range $0 \le z \le z_0$ (in what follows we
name this area in the extra space by core) we have nonzero source
functions $F_0 \left( z \right)$ and $F\left( z \right)$ which
correspond to the solution (\ref{internal1}) with decreasing scale
factor $\phi _{{\rm{int}}{\rm{.}}}$, and out of the core, i.e. in
the range $z
> z_0$ there are no source functions, i.e. $F_0 \left( z \right)\equiv F\left( z \right)\equiv
0$ if $z>z_{0}$. At present we do not argue the origin of such
matter distribution in the $6D$-space-time. Let us denote by
$\phi$ and $g$ the scale factors of solution to Einstein's
equations in this special case. It is obvious that they can be
constructed from the scale factors of solutions (\ref{internal1})
and (\ref{SolutionNoSource}) in these two regions in the following
way: $\phi \left( z \right) = \left[ \begin{array}{l}
\phi _{{\rm{int}}} \left( z \right),\ \ \ 0 \le z \le z_0 , \\
\phi _{{\rm{ext}}} \left( z \right),\ \ \ z > z_0 , \\
\end{array}\ \ ; \ \ \right.{\rm{      g}}\left( z \right) = \left[ \begin{array}{l}
g_{{\rm{int}}} \left( z \right),\ \ \ 0 \le z \le z_0 , \\
g_{{\rm{ext}}} \left( z \right),\ \ \ z > z_0 . \\
\end{array} \right.$
To avoid singularities we impose on the scale factors and there
first derivatives the continuity conditions  at the boundary of
the core $z=z_{0}$ : $\phi _{{\rm{int}}} \left( {z_0 } \right) =
\phi _{{\rm{ext}}} \left( {z_0 } \right);\phi '_{{\rm{int}}}
\left( {z_0 } \right) = \phi '_{{\rm{ext}}} \left( {z_0 }
\right);~g_{{\rm{int}}} \left( {z_0 } \right) = g_{{\rm{ext}}}
\left( {z_0 } \right);g'_{{\rm{int}}} \left( {z_0 } \right) =
g'_{{\rm{ext}}} \left( {z_0 } \right).$  So far there were the
following free parameters in the model: $M,\Lambda ,\delta
,d,\varepsilon ,p,a,z_0$. Solving the continuity conditions  we
can fix the following parameters $a = \frac{{2 - (x_0 - 1)\ln
x_0}}{{4x_0}}\lambda \left( x_0 \right),~~b=\frac{{\delta \Lambda
}}{{10M^4 }} = \frac{{x_0 - 1}}{{4x_0}}p\lambda \left( x_0
\right),~~d = \frac{{\left( {x_0 - 1} \right)e^{x_0} }}{{\lambda
\left( x_0 \right)}}$, where we have denoted $x_0=z_{0}^p>1$ and
$\lambda \left( x \right) = {\left( {x - 1} \right)e^x  + x + 1}
$. Now we have following free parameters in the model: $M, \delta,
\varepsilon ,p$ and $ z_0>1$.

Let turn our attention to the localization of the bulk
spin-${\raise0.7ex\hbox{$1$} \!\mathord{\left/
 {\vphantom {1 2}}\right.\kern-\nulldelimiterspace}
\!\lower0.7ex\hbox{$2$}}$ fermions on the brane described by our
solution. In this case equations for the components $\upsilon_l$
and $\zeta_l$ of the radial part $\xi _l \left( z \right)$ of
zero-mass wave-function have the same form as in the previous
section, and the solution reads
\begin{equation}
\xi _l \left( z \right) = \phi ^{ - 2} g^{ - \frac{1}{4}} z^{ -
\frac{1}{2}} \left( {\begin{array}{*{20}l}   {a_l z^l }  \\   {b_l z^{ - l} }  \\
\end{array}} \right) = \left[ {\begin{array}{*{20}l}   {\phi _{{\rm{int}}
}^{ - 2} g_{{\rm{int}}}^{ - \frac{1}{4}} z^{ - \frac{1}{2}}
\left( {\begin{array}{*{20}c}   {a_l z^l }  \\   {b_l z^{ - l} }  \\
\end{array}} \right),~~0 \le z \le z_0 ,}  \\
{ \phi _{{\rm{ext}}}^{ - 2} g_{{\rm{ext}}}^{ - \frac{1}{4}} z^{ -
\frac{1}{2}} \left( {\begin{array}{*{20}c}
{a_l z^l }  \\  {b_l z^{ - l} }  \\
\end{array}} \right),~~z > z_0 ,}  \\
\end{array}} \right.
\end{equation} with the following normalization requirement
\begin{equation}\label{NormalizationRequirement}1 = 2\pi \varepsilon
\left[ {\sqrt {\delta dp} \int\limits_0^{z_0 } {z^{\frac{p}{2} -
1} e^{ - \frac{{z^p }}{2}} \left( {a_l^2 z^{2l}  + b_l^2 z^{ - 2l}
} \right)dz}  + \sqrt {\delta b} \int\limits_{z_0 }^{ + \infty }
{\frac{{a_l^2 z^{2l}  + b_l^2 z^{ - 2l} }}{{z\left( {a + b\ln z}
\right)}}dz} } \right].\end{equation} Analysis of convergence of
the last integral at the boundaries of the whole integration
domain (i.e., at the points $z=0$ and $z=+\infty$)  shows  that
normalizable components of the zero-mass mode exist only in the
following cases: I) $a_l \ne 0~,~ b_l=0~,~ -p/4<l<0~$;   II) $a_l
= 0~,~ b_l\ne 0~,~ 0<l<p/4~$. So the number of normalizable
zero-mass modes depends on the value of parameter $p$. Namely in
the case $12 < p \le 16$ we have exactly three normalizable
zero-mass solutions with $ l = \pm1, \pm2, \pm3$. In this case the
effective wave-functions of this zero-mass modes in flat space
are:
\begin{equation}\label{WaveFunctionFlatSpace2}\Psi \left( {x^M } \right) = \left[ {\begin{array}{*{20}c}
{\psi \left( {x^\mu  } \right) \otimes \left( {\delta
dp\varepsilon ^2 } \right)^{\frac{1}{4}} z^{\frac{p}{2} - 1} e^{ -
\frac{{z^p }}{4}} \sum\limits_{0 < l < \frac{p}{4}} {\left(
{\begin{array}{*{20}c}
{a_{ - l} z^{ - l} e^{ - il\theta } }  \\
{b_l z^{ - l} e^{il\theta } }  \\
\end{array}} \right)} ,0 \le z \le z_0 ,}  \\
{\psi \left( {x^\mu  } \right) \otimes \left( {\delta b\varepsilon
^2 } \right)^{\frac{1}{4}} z^{ - 1} \left( {a + b\ln z} \right)^{
- 1} \sum\limits_{0 < l < \frac{p}{4}} {\left(
{\begin{array}{*{20}c}
{a_{ - l} z^{ - l} e^{ - il\theta } }  \\
{b_l z^{ - l} e^{il\theta } }  \\
\end{array}} \right)} ,z > z_0 .}  \\
\end{array}} \right.\end{equation}
They are localized in the radial direction of the extra space at
the following points \ \ $r_{{\rm{ferm}}{\rm{,}}l =   1}  =
\varepsilon \left( {1 - {6 \mathord{\left/ {\vphantom {6 p}}
\right. \kern-\nulldelimiterspace} p}} \right)^{{1 \mathord{\left/
{\vphantom {1 p}} \right. \kern-\nulldelimiterspace} p}} ,\ \
r_{{\rm{ferm}}{\rm{,}}l =  2}  = \varepsilon \left( {1 - {{10}
\mathord{\left/ {\vphantom {{10} p}} \right.
\kern-\nulldelimiterspace} p}} \right)^{{1 \mathord{\left/
{\vphantom {1 p}} \right. \kern-\nulldelimiterspace} p}} ,\ \
r_{{\rm{ferm}}{\rm{,}}l =   3}  = \varepsilon \left( {1 - {{14}
\mathord{\left/ {\vphantom
{{14}p}}\right.\kern-\nulldelimiterspace} p}} \right)^{{1
\mathord{\left/ {\vphantom {1 p}} \right.
\kern-\nulldelimiterspace} p}}$ . It must be mentioned that if
$12<b<14$ the last formula does not work, but in this subcase
directly from the wave function (\ref{WaveFunctionFlatSpace2}) we
have $r_{{\rm{ferm}}{\rm{,}}l = 3}  = 0$. For the boundary of the
core we have $z_{0}>1$, so each of these three modes are sharply
localized inside the core  and the wave-functions of these
localized modes have the "Gaussian shapes" in the extra space.
These three components of a higher dimensional fermion (stuck at
different points on the brane in the extra space) can be
identified with different $4$-dimensional fermion generations, and
so we have purely gravitational mechanism explaining the origin of
three generations of the Standard Model fermions.
\section{Non-zero-mass modes}
Now let us consider the case $m\ne0$ for the physical situation
described in the previous section. We have to solve the equations
(\ref{FermionEquationZeroModesGeneral}) in two regions - inside
the core and outside the core. We will seek the solution in the
form $\upsilon _l \left( z \right) = \phi ^{ - 2} g^{ -
\frac{1}{4}} z^{l - \frac{1}{2}} A\left( z \right)~,~\zeta _l
\left( z \right) = \phi ^{ - 2} g^{ - \frac{1}{4}} z^{ - l -
\frac{1}{2}} B\left( z \right),$ where $\phi=\phi_{{\rm{int}}}$
and $g=g_{{\rm{int}}}$ if $0\le z\le z_{0}$, and
$\phi=\phi_{{\rm{ext}}}~,~g=g_{{\rm{ext}}}~$ if $ z> z_{0}$. From
(\ref{FermionEquationZeroModesGeneral}) for the functions $A$ and
$B$ we get equations $\frac{d}{{dz}}\left[ {\frac{\phi }{{\sqrt g
}}z^{2l} \frac{{dA}}{{dz}}} \right] - m^2 \varepsilon ^2
\frac{{\sqrt g }}{\phi }z^{2l} A = 0~,~\frac{d}{{dz}}\left[
{\frac{\phi }{{\sqrt g }}z^{ - 2l} \frac{{dB}}{{dz}}} \right] -
m^2 \varepsilon ^2 \frac{{\sqrt g }}{\phi }z^{ - 2l} B = 0~$. We
examine the solution $A(z)$, the results for the $B(z)$ are the
same. Taking into account (\ref{SolutionNoSource}) and
normalization requirement for wavefunction at the radial infinity
in the extra space, outside the core the equation for
$A_{{\rm{ext}}}$ and its nomalizable solution assume the following
form: $x^2 A_{{\rm{ext}}}'' + \left( {\frac{2l}{p} + 1}
\right)xA_{{\rm{ext}}}' - \frac{{m^2 \delta b}}{{p^2
}}A_{{\rm{ext}}} = 0~,~~A_{{\rm{ext}}}\left( x \right) = \alpha
x^{ - \frac{{l + \sqrt {l^2  + m^2 \delta b} }}{p}}$, where
$\alpha$ is the integration constant and we have introduced new
variable $x=z^p$. Inside the core the equation for
$A_{{\rm{int}}}$ has the form
\begin{equation}\label{SolutionInside the Core}
\frac{d}{{dx}}\left[ {\sigma \left( x \right)x^{\frac{{4l +
p}}{{2p}}} \frac{{dA_{{\rm{int}}} }}{{dx}}} \right] - \frac{{m^2
\delta d}}{p}\frac{{x^{\frac{{4l - p}}{{2p}}} }}{{\sigma \left( x
\right)}}A_{{\rm{int}}}  = 0,~~\rm{where} ~~\sigma \left( x
\right) = \left[ {1 - d\left( {1 - e^{ - x} } \right)}
\right]e^{\frac{x}{2}}.\end{equation} At the origin of extra space
on the solution we impose condition $\left| {A_{{\rm{int}}} \left(
0 \right)} \right| <  + \infty$, and at the boundary of the core
$x=x_0$ we require the following matching conditions:
$A_{{\rm{int}}} \left( {x_0 } \right) = A_{{\rm{ext}}} \left( {x_0
} \right)~,~~A'_{{\rm{int}}} \left( {x_0 } \right) =
A'_{{\rm{ext}}} \left( {x_0 } \right)$. So in the core, $0\le x
\le x_0$, we have a Sturm-Liouville differential equation for
$A_{\rm{int}}$ with the following boundary conditions: $\left|
{A_{{\rm{int}}} \left( 0 \right)} \right| <  +
\infty~,~~\frac{{A'_{{\rm{int}}} \left( {x_0 } \right)}}{{A\left(
{x_0 } \right)}} =  - \frac{{l + \sqrt {l^2  + m^2 \delta b}
}}{{px_0 }}$. In the vicinity of the extra space origin for the
asymptotic form of (\ref{SolutionInside the Core}) and for its
normalizable solution we have: $xA'' + \left( {2\frac{l}{p} +
\frac{1}{2}} \right)A' - \frac{{m^2 \delta d}}{p}A =
0~,~~A_{{\rm{int}}} \left( x \right) = \beta x^{ - \frac{{4l -
p}}{{4p}}} BesselJ\left( {\nu ,\omega \sqrt x } \right)~,$ where
$\nu  = \frac{{4l - p}}{{2p}}~,~~\omega  = 2\sqrt { \frac{{m^2
\delta d}}{p}}~$ and $\beta$ is the integration constant. To make
some interesting qualitative conclusions we consider this solution
as the correct solution on the entire core range and choose $x_0$
sufficiently large to use the asymptotic expression for the Bessel
functions ($\left. {BesselJ(\nu ,y)} \right|_{y \to + \infty }
\sim \sqrt {\frac{2}{{\pi y}}} \cos \left( {y - \frac{{\nu \pi
}}{2} - \frac{\pi }{4}} \right)$) on the boundary of the core .
Then for the eigenvalue $\Delta$ we have the following
transcendental equation: $\tan \left( {2\Delta  - \frac{{\pi
l}}{p}} \right) =\Delta^{-1}{\sqrt {\frac{{l^2 }}{{p^2 }} +
\frac{b}{{pdx_0 }}\Delta ^2 } } ~,~~\Delta  = \sqrt {\frac{{m^2
\delta dx_0 }}{p}}~.$ This equation gives the discrete spectrum
for $\Delta=\Delta_1, \Delta_2, \Delta_3, ...$ , and for
corresponding nonzero masses we get: $m_n  = \sqrt
{\frac{p}{{\delta dx_0 }}} ~\Delta _n$. Between zero and nonzero
modes there is the gap depending on the value of parameter
$\delta$, so choosing the appropriate value of $\delta$ the
non-zero modes can be neglected at low energies. The same
conclusions can be done with respect to the exact solution of
(\ref{SolutionInside the Core}), although it is difficult to find
the solutions explicitly.
\section{Conclusion} To conclude we summarize our main results.
In this article in the case of $(1+5)$-space-time we have
introduced new brane-solutions with exponential scale factors
sharply localizing the zero modes of all local fields (spin $0$
scalar field, spin ${\raise0.7ex\hbox{$1$} \!\mathord{\left/
 {\vphantom {1 2}}\right.\kern-\nulldelimiterspace}
\!\lower0.7ex\hbox{$2$}}$ spinor field, spin 1 gauge field, spin
${\raise0.7ex\hbox{$3$} \!\mathord{\left/
 {\vphantom {3 2}}\right.\kern-\nulldelimiterspace}
\!\lower0.7ex\hbox{$2$}}$ gravitino field, spin $2$ gravitational
field and totally antisymmetric tensor fields) on the brane in the
extra space. We have explicitly shown that these solutions
localize infinite number of $4D$-fermions on the brane. To solve
this problem we have introduced another specific brane-solution
with realistic source functions, i.e. the source functions $F_{0}$
and $F$ are nonzero inside the core $0\le z \le z_{0}$ in the
extra space, and $F_{0} \equiv F \equiv 0$ outside the core. We
also have explicitly shown that in this model there are cases when
exactly three $4D$-fermion generations can naturally arise only
through gravity. As the zero modes of all other local bulk fields,
these three zero modes of bulk fermion are sharply peaked at
different positions on the brane and have "Gaussian shape"
wave-functions in the extra space. So there is possibility to
provide a framework for understanding the fermion mass hierarchy
in terms of higher dimensional geography \cite{AH-Schmaltz}.

\acknowledgments Author would like to acknowledge the hospitality
extended during his visits at the Abdus Salam ICTP, and to thank
Irma Razmiashvili for typing of the article.

\end{document}